\documentclass[prd,twocolumn,superscriptaddress,nofootinbib,notitlepage]{revtex4-1} 
\usepackage[utf8]{inputenc}
\usepackage{hyperref}
\usepackage{amsfonts}
\usepackage{enumitem}
\hypersetup{
    colorlinks=true,
    linkcolor=blue,
    filecolor=magenta,      
    urlcolor=cyan,
    pdfpagemode=FullScreen,
}
\usepackage{verbatim}
\usepackage{upgreek}
\usepackage[normalem]{ulem}
\usepackage{graphicx}
\usepackage[dvipsnames]{xcolor}
\usepackage{soul}
\usepackage{ulem, color, framed}
\usepackage{amsmath}
\usepackage{subfigure}
\usepackage{braket}

\begin{document}
\title{Accelerated Quantum Circuit Monte-Carlo Simulation for Heavy Quark Thermalization}
\author{Xiaojian Du}
\email[]{xiaojian.du@usc.es}
\affiliation{Instituto Galego de Fisica de Altas Enerxias (IGFAE), Universidade de Santiago de Compostela, E-15782 Galicia, Spain}
\author{Wenyang Qian}
\email[]{qian.wenyang@usc.es}
\affiliation{Instituto Galego de Fisica de Altas Enerxias (IGFAE), Universidade de Santiago de Compostela, E-15782 Galicia, Spain}
\date{\today}

\begin{abstract}

Thermalization of heavy quarks in the quark-gluon plasma (QGP) is one of the most promising phenomena for understanding the strong interaction. The energy loss and momentum broadening at low momentum can be well described by a stochastic process with drag and diffusion terms.
Recent advances in quantum computing, in particular quantum amplitude estimation (QAE), promise to provide a quadratic speed-up in simulating stochastic processes. 
We introduce and formalize an accelerated quantum circuit Monte-Carlo (aQCMC) framework to simulate heavy quark thermalization. 
With simplified drag and diffusion coefficients connected by Einstein's relation, we simulate the thermalization of a heavy quark in isotropic and anisotropic mediums using an ideal quantum simulator and compare that to thermal expectations. With Grover-like QAE, we calculate physical observables with quadratically fewer resources, which is a boost over the classical MC simulation that usually requires a large sampling number at the same estimation accuracy.
\end{abstract}

\maketitle

\section{Introduction}

Thermalization is one of the most important common features of a non-equilibrium system. An open system that undergoes quantum decoherence by rapidly exchanging information with the environment usually tends to thermalize conventionally and classically. Heavy quark thermalization in the background of quark-gluon plasma (QGP) produced in relativistic heavy-ion collisions (HICs) is such an open system that heavy quarks have distinguished separation of scales compared to the soft QGP medium.
With the shear viscosity over entropy density ratio $\eta/s$ characterizing the speed of hydrodynamization and temperature $T$ as the only scale, the thermalization time of the QGP is characterized by a scale of $\tau_h\simeq 4\pi\eta/(Ts)$~\cite{Kovtun:2004de}. Compared to the soft QGP, the relaxation of the heavy quark is prolonged by its heavy mass $\tau_R\simeq m_{\rm HQ}\tau_h/T$. With a typical charm quark mass of $\simeq$1.5\,GeV, and the QGP medium temperature of 300-500\,GeV in HICs, heavy quark undergoes a time of thermalization caused and dominated by a thermal environment. This is more extreme for the bottom quark with $\simeq$ 4.5\,GeV mass, that the thermalization process is not even finished in a 10\,fm of the QGP phase. Eventually, the hadronized heavy flavors measured in the detector are not thermalized and the characterization of the heavy quark spectra tells us the medium property of the QGP. 

The heavy masses of heavy quarks not only delay the thermalization in the QGP medium but also make the heavy quarks less relativistic compared to the almost massless partons in the QGP. This leads to a well-established thermalization description for heavy quarks based on a stochastic process with low-momentum random kicks from the medium~\cite{Moore:2004tg,Cao:2011et,He:2011qa,Das:2013kea,Adare:2013wca,Ke:2018tsh,Du:2022uvj,Pooja:2023gqt,Guo:2023phd,Singh:2023smw}. The thermalization in this description is controlled by two competing effects, the energy loss from a drag term and a diffusion from a stochastic term. The energy loss tends to reduce the momentum of a heavy quark while the diffusion tends to broaden the momentum distribution.
The competing contributions eventually thermalize the heavy quark to a certain distribution controlled by a fluctuation-dissipation theorem, known as Einstin's relation. 
In a non-relativistic or static limit, the thermal distribution is given by a classical Maxwell-Boltzmann distribution. For more discussions on heavy quark thermalization in HIC phenomenology, see reviews~\cite{Rapp:2018qla,He:2022ywp,Zhao:2023nrz}. Notably, this stochastic process is so generic that it is not limited to the description of a heavy quark thermalization but is broadly utilized in many research topics, such as the Black–Scholes model in quantitative finance, which in part inspired our work. 

Quantum computing technology, using laws of quantum mechanics, has already been extensively applied in many areas of nuclear physics~\cite{Wiesner:1996xg, Jordan:2011ci, Jordan:2014tma, Preskill_2018, DeJong:2020riy,Kharzeev:2020kgc, Czajka:2021yll, Li:2021kcs, Ngairangbam:2021yma, Barata:2022wim, Bauer:2022hpo, Gustafson:2022dsq, Barata:2023, Brown:2023llg}, where the strength of quantum computing is usually exploited from its exponential state space, local Hamiltonian simulation, and near-term variational algorithms. Recently, novel gate-based quantum finance strategy~\cite{Stamatopoulos:2020xez, Woerner_2019, Rebentrost:2018, montanaro2015quantum} with the quantum amplitude estimation (QAE)~\cite{brassard2002quantum} exhibits a promising quadratic speed-up over the classical Monte-Carlo (MC) method. In much the same spirit as Grover's algorithm~\cite{Grover:1996, Grover:1997}, the QAE allows efficient estimation of the amplitude of the designated quantum state.
The main contribution of this work is the first application of an accelerated quantum circuit Monte-Carlo (aQCMC) strategy using the QAE techniques for heavy-quark thermalization. 
Different events are simulated as quantum state evolution with sufficient quantum shots, and the physical observables are efficiently extracted with amplitude estimation. 
With the constant improvements in the QAE algorithms~\cite{grinko2021iterative, suzuki2020amplitude}, the aQCMC may expect to become a more standard approach, especially in future large-scale quantum simulations. 

This manuscript is organized as follows. In Sec.~\ref{sec:heavyquark}, we review the heavy-quark thermalization formulated as a stochastic differential equation and its standard classical simulation strategy with the MC method. In Sec.~\ref{sec:quantum}, we discuss the aQCMC strategy utilized in this work to speed up the computation. In Sec.~\ref{sec:results}, we present our simulation results in isotropic and anisotropic mediums using {\tt Qiskit}. In Sec.~\ref{sec:conclusion}, we summarize and discuss future avenues of this work.

\section{Heavy quark thermalization}\label{sec:heavyquark}

\subsection{Stochastic description of heavy quark thermalization}
The heavy-quark thermalization can be characterized by a stochastic differential equation (SDE) known as the Langevin equation~\cite{Moore:2004tg}, evolving in full position-momentum phase space $\vec{x},\vec{p}$ with evolution time $t$
\begin{align}
\nonumber
dx_i&=\frac{p_i}{E(\vec{p})}dt,~~~i=x,y,z,\\
dp_i&=-A(\vec{x},\vec{p},t)p_{i}dt+\sigma_{ij}(\vec{x},\vec{p},t)dW_j,
\label{eq-langevin}
\end{align}
where the random force that sampled as a Wiener process $dW\sim\mathcal{N}(0,dt)$ has correlation $\braket{dW_idW_j}=\delta_{ij}dt$.
The drag coefficient $A(\vec{x},\vec{p},t)$ and the diffusion coefficient $\sigma_{ij}(\vec{x},\vec{p},t)$ in HICs may be calculated from either quantum chromodynamics (QCD)~\cite{Svetitsky:1987gq,Romatschke:2004au,Mustafa:2004dr,Caron-Huot:2007rwy,Liu:2020dlt,Altenkort:2023eav,Du:2023izb,Scheihing-Hitschfeld:2023tuz,Dang:2023tmb}, or QCD-like theories~\cite{Herzog:2006gh,Gubser:2006qh,Casalderrey-Solana:2006fio,Herzog:2006gh,Akamatsu:2023hwi,Pandey:2023dzz,Avramescu:2023qvv,Boguslavski:2023fdm,Boguslavski:2023jvg} with a heavy quark interacting with the medium.
Applying Ito's lemma, the Langevin equation Eq.~(\ref{eq-langevin}) can be reformulated as a Kolmogorov-forward equation, known as the Fokker-Planck equation, presenting the time evolution of the heavy quark non-equilibrium distribution $f(\vec{x},\vec{p},t)$ as
\begin{align}
\nonumber
\frac{\partial }{\partial t}f(\vec{x},\vec{p},t)&=\frac{\partial }{\partial p_i}\left[A(\vec{x},\vec{p},t)p_if(\vec{x},\vec{p},t)\right]\\
&+\frac{\partial^2 }{\partial p_i\partial p_j}\left[B_{ij}(\vec{x},\vec{p},t)f(\vec{x},\vec{p},t)\right],
\label{eq-fokkerplank}
\end{align}
with the diffusion coefficient $B_{ij}(\vec{x},\vec{p},t)=\sigma_{ik}(\vec{x},\vec{p},t)\sigma_{jk}(\vec{x},\vec{p},t)/2$.
There is no general solution to the Fokker-Planck equation Eq.~(\ref{eq-fokkerplank}) and the evolution would depend on the initial condition. 
However, the solution to the Fokker-Planck equation would be an attractor towards the thermal limit.
This transition from various ordered initial conditions to a unique chaotic limit is the thermalization of heavy quarks within a medium. 
These transport coefficients are generally medium profile dependent, but in a thermal and homogeneous medium, we may drop the spatial $\vec{x}$ and time $t$ dependencies. 
The perturbative QCD calculation suggests the drag coefficient $A(\vec{p})$ to be almost a constant at low momentum $p\lesssim 2M$~\cite{Du:2023izb}.

With an approximately constant drag coefficient, one may simplify the Langevin equation in the non-relativistic limit at a small momentum $p$, which may be further rescaled by the heavy quark mass $M$.
Keeping diagonal terms only in the diffusion term, these simplifications lead to a dimensionless Langevin equation
\begin{align}
dq_i=-q_id\tilde{t}+d\tilde{W}_i.
\label{eq-langevin_unity}
\end{align}
In the above equation we have used dimensionless momentum $q_i=p_i/M$, time $d\tilde{t}=Adt$, and anisotropic stochastic terms $d\tilde{W}_i\sim\mathcal{N}(0,2Td\tilde{t}/(M\chi_i^2))$ with proper Einstein's relation $A=\sigma_{ii}^2\chi_{i}^2/(2MT)$. See App.~\ref{app:nr_limit} for details of the derivations, and also Refs~\cite{Hauksson:2021okc,Barata:2022utc,Boguslavski:2023fdm,Boguslavski:2023jvg,Du:2023izb,Prakash:2023wbs} for recent discussions on hard probes in anisotropic medium.
Notice that the heavy quark relaxation time $\tau_{\rm R}\simeq 1/A$, the value of $d\tilde{t}$ represents the speed of energy loss and thermalization.
Thus, a realistic simulation would favor the $d\tilde{t}$ to be as small as possible, and a value of $d\tilde{t}\simeq 1/N_t$ takes about $N_t$ steps to thermalize (thermalization will also be delayed by a large momentum, for instance, a heavy quark jet).
Another relevant scale is the temperature over heavy quark mass ratio $T/M$ in the variance $\tilde{\sigma}_i^2d\tilde{t}=2Td\tilde{t}/(M\chi_i^2)$.
The dimensionless Fokker-Planck equation corresponding to Eq.~(\ref{eq-langevin_unity}) reads
\begin{align}
\frac{\partial }{\partial \tilde{t}}f(\vec{q},\tilde{t})&=\frac{\partial }{\partial q_i}\left[q_if(\vec{q},\tilde{t})\right]
+\frac{1}{2}\tilde{\sigma}_{ii}^2\frac{\partial^2 }{\partial q_i^2}\left[f(\vec{q},\tilde{t})\right]
\label{eq-fokkerplank-unity}
\end{align}
The thermal distribution in terms of these dimensionless quantities reads 
\begin{eqnarray}\label{eq:feq_no_unit}
f^{\rm eq}(\vec{q})\propto\exp\bigg[-\frac{q_x^2}{\tilde{\sigma}_x^2}-\frac{q_y^2}{\tilde{\sigma}_y^2}-\frac{q_z^2}{\tilde{\sigma}_z^2}\bigg]
\end{eqnarray}

This stochastic process is usually simulated with the MC methods, by sampling the Wiener process for each time step. The trajectory of a heavy quark contributes to an event and a collection of these events provides a time series of the heavy quark distribution towards thermalization.
On a modern digital computer, this MC simulation is straightforward: one starts with whatever heavy quark initial distribution $f(\vec{q},\tilde{t}_0)$, and samples the heavy quark initial momentum $(q_x^{\tilde{t}_0},q_y^{\tilde{t}_0},q_z^{\tilde{t}_0})$ accordingly. Similarly, the values of the stochastic variables $(dW_x^{\tilde{t}},dW_y^{\tilde{t}},dW_z^{\tilde{t}})$ can be uncorrelatedly sampled with a set of independent normal distributions 
$\{\mathcal{N}(0,2Td\tilde{t}/(M\chi_i^2))\}$ with $i=x, y, z$ for each time in a diagonalized form. The increment of the momentum follows the Langevin equation Eq.~(\ref{eq-langevin_unity}) and the momentum at the next step can be calculated with the forward-Euler method as
\begin{align}\label{eq:quantum-evolution}
q_i^{\tilde{t}+d\tilde{t}}=q_i^{\tilde{t}}-q_i^{\tilde{t}}d\tilde{t}+d\tilde{W}_{i}^{\tilde{t}},
\end{align}
Iterating the above algorithm for large enough $N_t$ steps from $\tilde{t}_0$ to $\tilde{t}_0+N_t d\tilde{t}$ gives a time series of one heavy quark momentum
\begin{align}
\mathbf{Q}^{T}=\{q_i^{\tilde{t}_0},q_i^{\tilde{t}_0+d\tilde{t}},\cdots,q_i^{\tilde{t}_0+N_t d\tilde{t}}\},
\end{align}
and repeating for a total of $N_\mathrm{event}$ events produces an emergent phenomenon of heavy quark thermalization, which leads to a thermal distribution 
\begin{align}
\{ \mathbf{Q}_{1}^{\tilde{t}_0+N_td\tilde{t}},\mathbf{Q}_{2}^{\tilde{t}_0+N_td\tilde{t}},\cdots,\mathbf{Q}_{N_{\rm event}}^{\tilde{t}_0+N_td\tilde{t}}\}\sim f^{\rm eq}(\vec{q}),
\end{align}
with $N_t d\tilde{t}\gg 1$. Then, for any physical quantity $F(\vec{q})$ at time $\tilde{t}$, its expectation value would be 
\begin{equation}
\braket{F(\vec{q})}=\frac{1}{N_{\rm event}}\sum_{i=1}^{N_{\rm event}} F(\mathbf{Q}_{i}^{\tilde{t}}).
\end{equation}
The MC simulation on a modern computer is straightforward but often requires large computational resources for reasonable precision. By encoding the stochastic process on the quantum circuit and accelerating with the QAE algorithms, one may reduce the inherent problem complexity faced in classical simulations, and obtain a quadratic quantum speed-up compared to the classical method to the same precision.

\section{Quantum strategy}\label{sec:quantum}

In this section, we formulate the quantum strategy, the quantum circuit Monte-Carlo (QCMC) to simulate the heavy quark thermalization in a stochastic description. For the QCMC simulation, we encode the particle's momenta $q_i$ in each direction as a quantum state. With a generic $n$-qubit quantum register, one has in principle $N=2^n$ possible modes for the heavy quark momenta $q$. By restricting the momentum $q\in[-q_{\rm max},q_{\rm max})$, we discretize $q$ into $N$ values with $\delta q=2q_{\rm max}/N$. Then, we further shift the physical momentum $q$ to the positive momentum $\bar{q}$ by a constant $q_\mathrm{max}$ so that $\bar {q}=q+q_{\rm max}\in[0,2q_{\rm max})$ and impose a periodic boundary condition, i.e. $\bar {q}= \bar {q} \text{ mod } (2q_\mathrm{max})$. The use of non-negative dimensionless momenta $\bar {q}$ makes a straightforward binary mapping onto the corresponding quantum states, which can be extended to all three spatial dimensions $x,y$, and $z$. 

\begin{figure}
    \centering
    \subfigure[\label{fig:QMCwithReset} The depth-oriented QCMC with resets]{
    \includegraphics[width=0.42\textwidth]{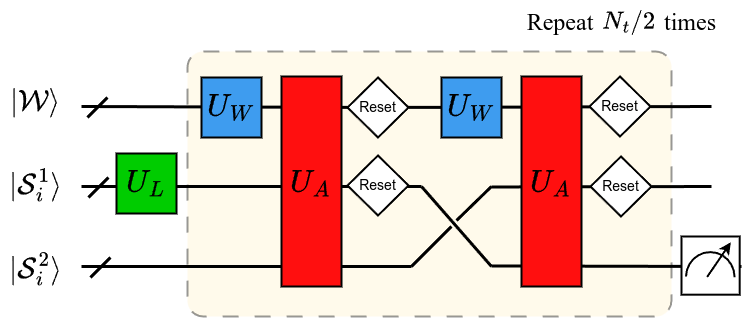}}
    \subfigure[\label{fig:QMCwithQAE} The breadth-oriented aQCMC with the QAE]{
    \includegraphics[width=0.46\textwidth]{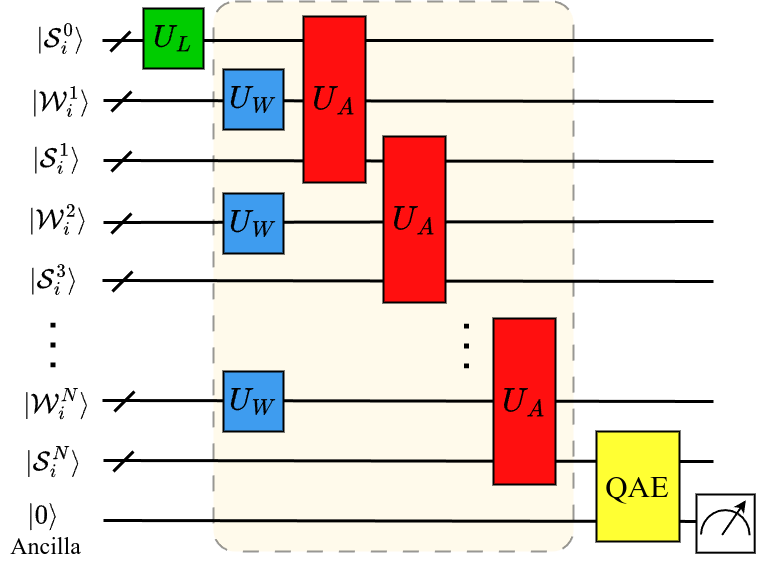}}
    \caption{The schematic quantum circuits of the QCMC and the aQCMC involving $N_t$ time steps in direction $i$ constructed using $\mathcal{S}$ and $\mathcal{W}$ registers (a) with {\tt reset} gates, and (b) without {\tt reset} gates but with the QAE. Quantum gates $U_L$, $U_W$, and $U_A$ are used to represent the initial distribution loading, stochastic diffusion, and quantum evolution, respectively. 
    \label{fig-circuit-schematic}}
\end{figure}

To thermalize with approximately $N_t$ steps, reasonable values of the coefficients for the simulation scale as $d\tilde{t}\simeq 1/N_t$ with $N_t>1$. The variance in the stochastic term is chosen to be $\tilde{\sigma}_i^2d\tilde{t}=2Td\tilde{t}/(M\chi_i^2)\simeq d\tilde{t}/(2\chi_i^2)$ according to scales of heavy quark mass $M$ and temperature $T$ in HICs. 
Since generic quantum multiplication and divisions are complicated~\cite{ruiz2017quantum}, we pick $d\tilde{t}=1/2^d$ with positive integer $d$ in practical simulations, which can be realized on the quantum circuit by shifting the quantum state with $d$ qubits using a sequence of $\mathsf{CX}$ gates. 

In general, for each of the $i = x,y,z$ directions, we prepare quantum register $\mathcal{S}_i$ to encode the particle's momentum $q_i$ and quantum register $\mathcal{W}_i$ to encode the diffusion term $d\tilde{W}_i$. Each register is represented by a set of qubits. The numbers of qubits $n_{\mathcal{S}},n_{\mathcal{W}}$ in registers $\mathcal{S}_i,\mathcal{W}_i$ are not necessarily the same. The increment at each time step $\tilde{t} = n d\tilde{t}$ contributed from the drag term $-q_id\tilde{t}$ and the diffusion term $-d\tilde{W}_i$ are implemented as unitary quantum operators $U_{A^n_i}$ following Eq.~\eqref{eq:quantum-evolution} so that 
\begin{align}\label{eq:UA_gate}
\nonumber
U_{A_i^n} &\ket{d\tilde{W}_i^n}_{\mathcal{W}_i^n} \otimes
\ket{q^n_i}_{\mathcal{S}^n_i} \otimes
\ket{0}_{\mathcal{S}^{n+1}_i}  \\
\nonumber
=& \ket{d\tilde{W}_i^n}_{\mathcal{W}_i^n}  \otimes
\ket{q^n_i}_{\mathcal{S}^n_i} \otimes
\ket{q_i^n - q_i^nd\tilde{t}+d\tilde{W}_{i}^n}_{\mathcal{S}^{n+1}_i}\\
=&\ket{d\tilde{W}_i^n}_{\mathcal{W}_i^{n}}  \otimes
\ket{q^{n}_i}_{\mathcal{S}^{n}_i} \otimes
\ket{q^{n+1}_i}_{\mathcal{S}^{n+1}_i}.
\end{align}
Here, $U_{A_i^n} = U_{A}$ is time-independent with a constant drag coefficient $A$, though it is not required. Now we introduce the quantum gates used in the circuit:

\begin{enumerate}
    \item \textbf{Distribution loading gates ($U_L$)} are responsible for loading the initial momentum distribution on the quantum register $\mathcal{S}$ for the system. In principle, one can start with either a single momentum or any momentum distribution for the heavy quark and evolve it on the circuit. Here, we initialize an arbitrary single momentum each time using $\mathsf{X}$ gates. 
    
    \item \textbf{Stochastic Wiener gates ($U_W$)} provide the stochastic contribution to the quantum circuit for the Wiener process $dW$. Here, we sample normal distribution $\mathcal{N}(0,\sigma=2Td\tilde{t}/(M\chi_i^2))$ exactly, and subsequent circuit transpilation automatically builds the quantum gates for the distribution. In other words, $U_W \ket{0} = \sum_{\bar {q}} \sqrt{\mathcal{P}(\bar {q})}\ket{\bar {q}/\delta q}$ with probability $\mathcal{P}(\bar {q}) = (1/\sqrt{2\pi\sigma^2})\exp(-\bar {q}^2/(2\sigma^2))$.
    
    \item \textbf{Quantum evolution gates ($U_A$)} are the main building blocks of the QCMC, where we follow Eq.~\eqref{eq:UA_gate} to construct the evolution gates. Specifically, we implement and utilize the quantum adders and multipliers (see App.~\ref{app:quantum_gates} for a brief review) to build the stochastic Langevin evolution at each time step. One additional constant quantum adder is included to remedy the momenta from $q$ to non-negative $\bar {q}$ per each step. Notably, these quantum arithmetic gates correspond directly to the classical arithmetic operations, though one still needs to manually manipulate these operations at the quantum-register level for today's quantum computers. Since quantum Fourier transforms are innate to most arithmetic operations, it may be more efficient to use Fourier basis as the encoding basis to abbreviate consecutive operations. 
\end{enumerate}

In principle, one could simulate the MC process on the quantum circuits as efficiently as on a classical computer. Nevertheless, since at each time step, the Wiener process $dW$ needs to be uncorrelated and the quantum arithmetic operations are on the register level, the quantum circuit would require additional sets of registers $\mathcal{S}$ and $\mathcal{W}$ for each time iteration, making the total qubit number scales as $\mathcal{O}((2N_t-1)n)$ assuming $n_{\mathcal{Q}}=n_{\mathcal{W}}=n$. To circumvent this tower-like quantum circuit, one may include $\mathsf{reset}$ gates to economically reuse the quantum registers repeatedly for different time steps, as in Fig.~\ref{fig:QMCwithReset}, leading to only $\mathcal{O}(3n)$ qubits. 

The quantum circuit Monte-Carlo (QCMC) method can be accelerated by taking advantage of the quantum amplitude estimation~\cite{brassard2002quantum} (QAE), a generalized version of Grover's search algorithm~\cite{Grover:1996, Grover:1997}. 
Suppose an operator $A_F$ acts on $n+1$ qubits,
\begin{align}
    A_F \ket{\psi}_n\ket{0} = \sqrt{1-a}\ket{\psi_0^*}_{n+1} + \sqrt{a}\ket{\psi_1^*}_{n+1} = \ket{\psi^*}_{n+1},
\end{align}
where $\ket{\psi_0^*} = \ket{\psi_0}_n\ket{0}$, $\ket{\psi_1^*} = \ket{\psi_1}_n\ket{1}$, and $a\in[0,1]$ is the desired expectation of interests. Specifically, $a = \braket{\psi|_n F | \psi}_n$ is the expectation of any physical quantity $F$ in terms of heavy quark momentum, whose distribution is represented by the quantum state $\ket{\psi}_n$. Using Grover operator $\mathcal{Q} = A_F S_0 A_F^\dagger S_{\psi_1}$ with $S_x$ sign-flipping operator on the state $x$, the QAE allows for high-probability estimation of $a$ in $N_q$ queries of $A_F$ with error $\epsilon = \mathcal{O}(1/N_q)$. This represents a quadratic speed-up over classical MC~\cite{brassard2002quantum}. 
Intuitively, one can consider Grover's operator $\mathcal{Q}\sim R_{\psi^*} R_{\psi^*_0}$ as sequence of two reflection operators $R_{\psi^*}$, $R_{\psi^*_0}$. This operator successively reflects about the ``bad" state $\ket{\psi^*_0}_{n+1}$ and the ``mean" state $\ket{\psi^*}_{n+1}$, such that the amplitude of the ``good" state $\ket{\psi^*_1}_{n+1}$ is amplified.
See App.~\ref{app:qae_circuit} for more technical details. 
In principle, one may use the standard quantum phases estimation (QPE) with extra auxiliary qubits~\cite{nielsen_chuang_2010, brassard2002quantum} to retrieve the amplitude where the estimation success rate is quickly boosted close to unity.

In practice, the QAE approach is usually difficult for two reasons: Firstly, universal oracle implementation for the expectation function $F$ is nontrivial; secondly, the QPE, the key to extract amplitude, requires expensive auxiliary qubits and substantial multi-qubit gates~\cite{brassard2002quantum}. Fortunately, operators $U_F$ involving piecewise linear functions can be approximated via Taylor expansion and implemented using controlled $\mathsf{R_Y}$ gates~\cite{Woerner_2019, Gacon2020}, so we are capable of investigating momentum and absolute momentum expectation of the particle, i.e., $F(q) = q$ and $F(q) = |q|$.
Alternative loading methods to reduce the circuit complexity that one may consider include quantum generative adversarial networks~\cite{zoufal2019quantum} and approximate quantum compiling~\cite{Madden:2021dax}.

\begin{figure}
    \centering
    \subfigure[\label{fig:p_X} Momentum expectation $\braket{q}$]{
    \includegraphics[width=0.21\textwidth]{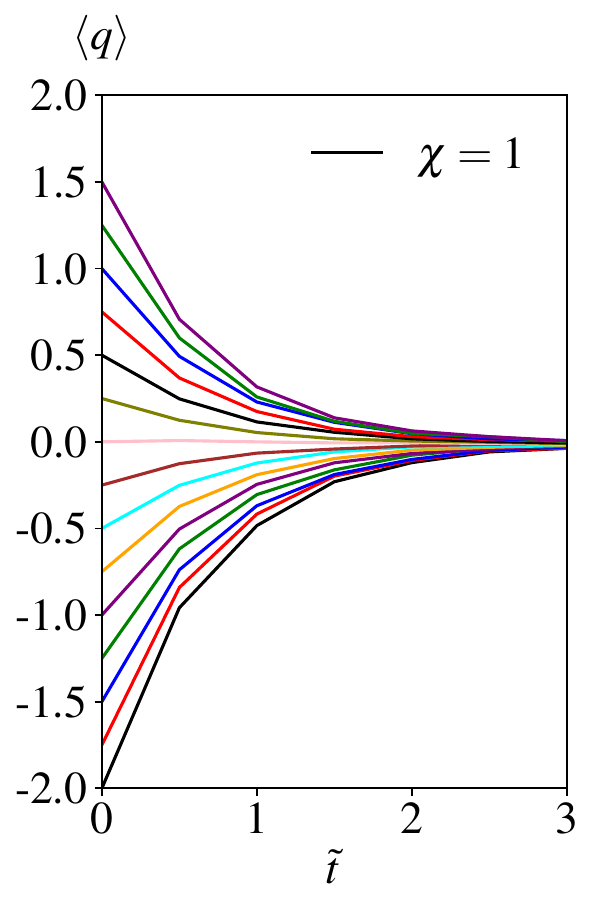}
    \includegraphics[width=0.21\textwidth]{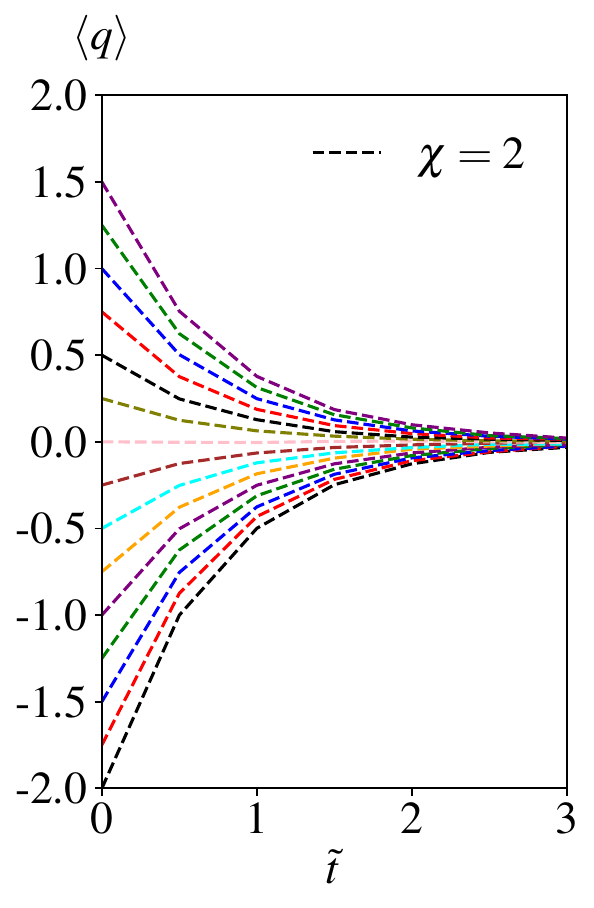}}
    \subfigure[\label{fig:Absp_X} Absolute momentum expectation $\braket{|q|}$]{
    \includegraphics[width=0.21\textwidth]{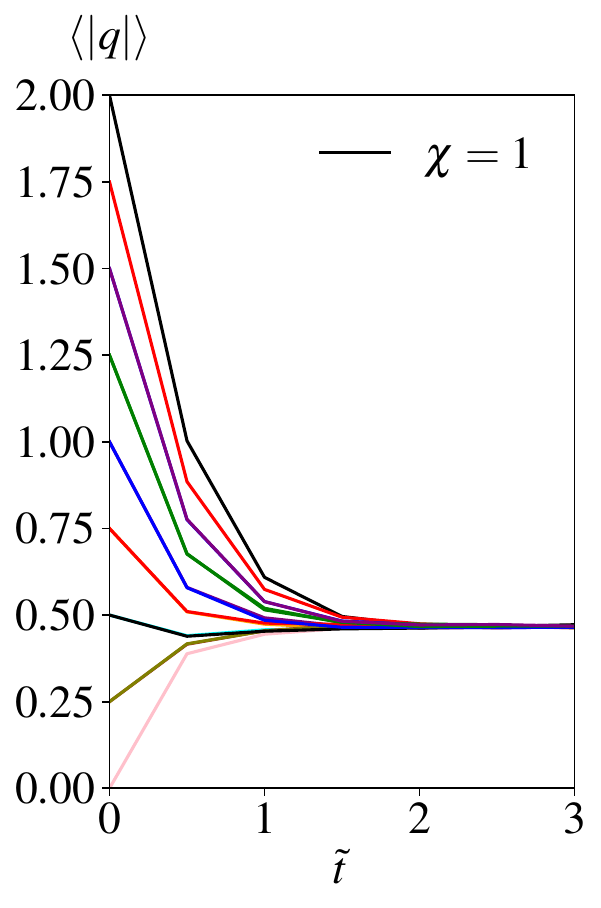}
    \includegraphics[width=0.21\textwidth]{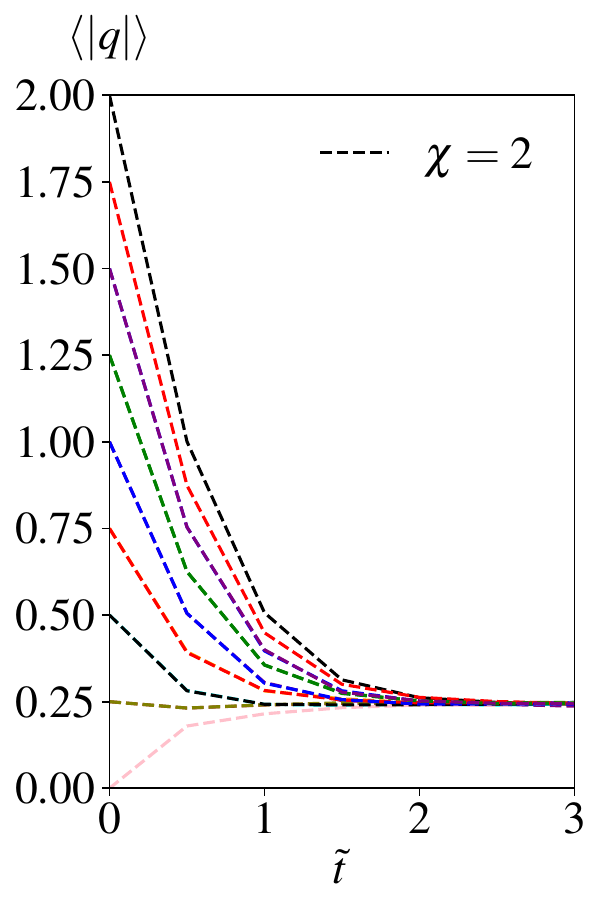}}
    \caption{\label{fig:qc_1D}Quantum simulation using the QCMC method in a one-dimensional system. Panels (a) and (b) show the expectations of $\langle q \rangle$ and $\langle |q| \rangle $ for differential initial momenta (marked in colored lines). 
    }
\end{figure}

On the other hand, the complexity of the QPE can be circumvented using novel QPE-free algorithms~\cite{grinko2021iterative, suzuki2020amplitude, Wie:2019, Nakaji_2020, Aaronson:2020, Manzano:2022keg, Manzano:2023eol}, which are mostly based on selected Grover iterations $Q^k A_F$ to estimate the quantum amplitude efficiently, and the same quadratic speed-up can be obtained~\cite{Aaronson:2020}. In particular, we focus on the Iterative QAE (IQAE) algorithm in our simulation result, which proves most economical in estimation accuracy and confidence level~\cite{grinko2021iterative} for our simulation resources. Alternatively, one may also consider using variational QAE with constant-depth circuits~\cite{Plekhanov2022variationalquantum}. Nonetheless, it is crucial to point out that by having Grover operators in the QAE we cannot use the non-unitary reset gates directly, and consequently, we regress to the tower-like quantum circuit in Fig.~\ref{fig:QMCwithQAE} when the QAE is involved.

\section{Simulation results }\label{sec:results}

\begin{figure}
    \centering
    \raisebox{0.2em}{\subfigure[\label{fig:QAE_p} Expectation $\braket{q}$]{
    \includegraphics[width=0.21\textwidth]{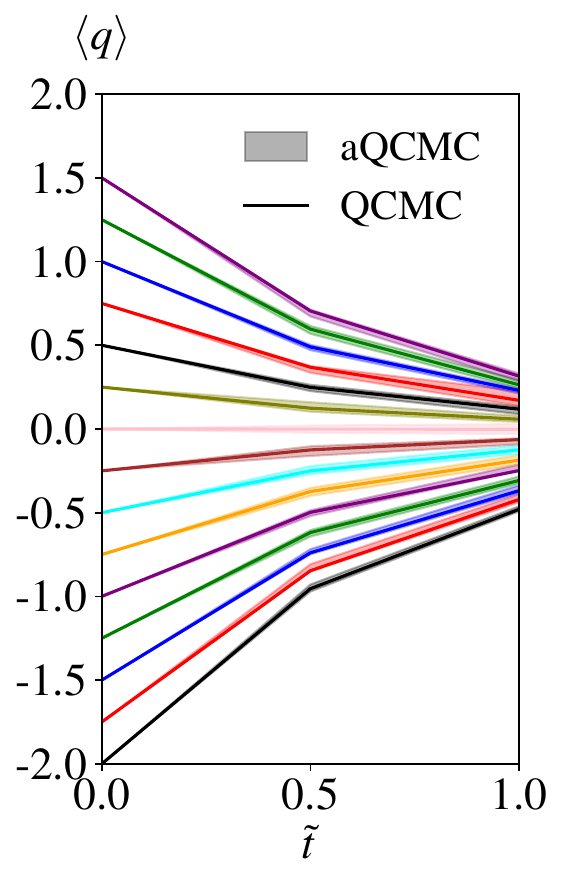}}}
    \subfigure[\label{fig:QAE_Absp} Expectation $\braket{|q|}$]{
    \includegraphics[width=0.215\textwidth]{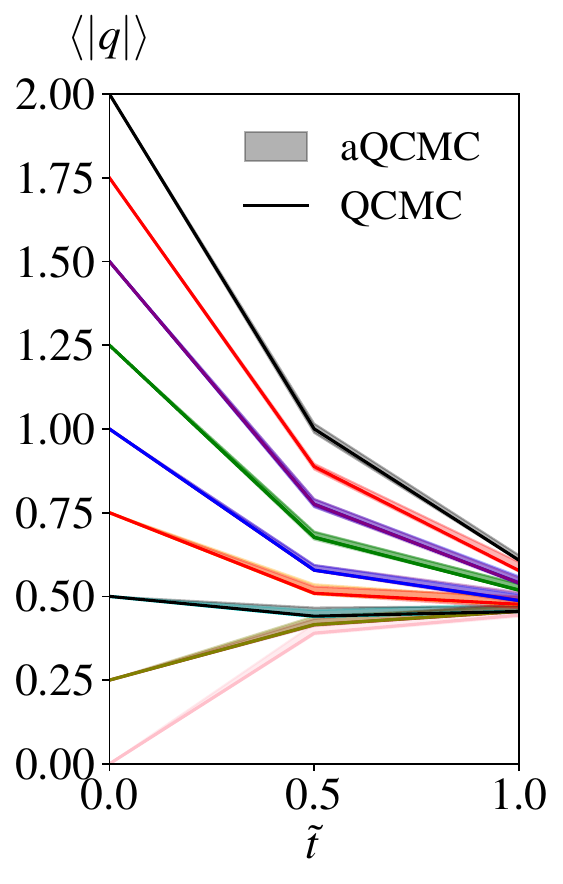}}
    \caption{\label{fig:qc_1D_QAE}
    Quantum simulation using the aQCMC with Iterative QAE~\cite{grinko2021iterative} (in shaded area) compared to QCMC with direct measurement (in lines) for the earlier time steps.
    }
\end{figure}

\begin{figure}
    \centering
    \includegraphics[width=0.45\textwidth]{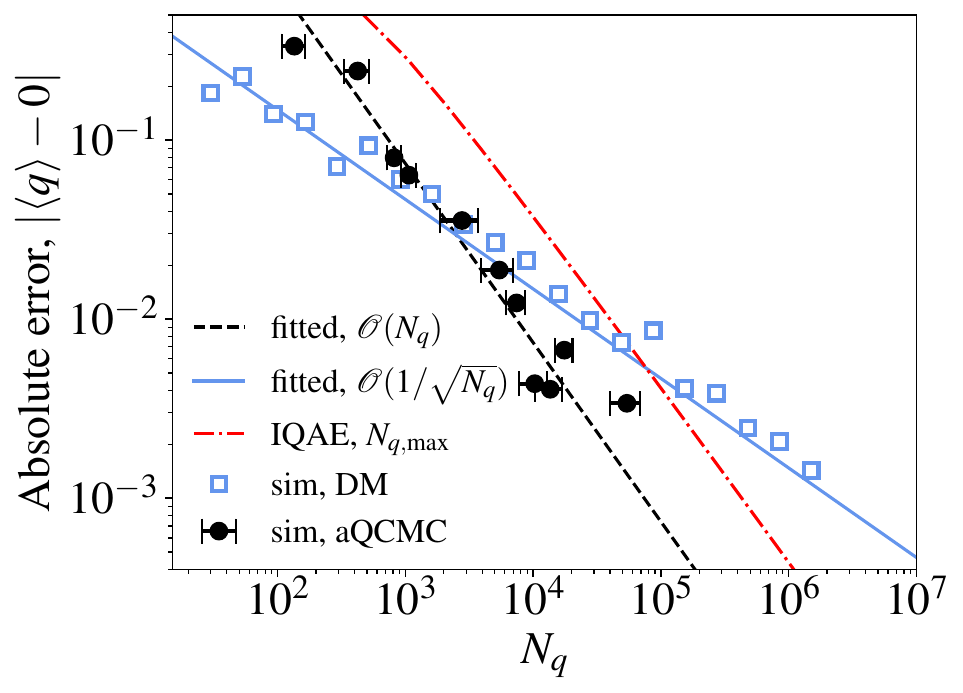}
    \caption{\label{fig:quantum_advantage}
    Quantum advantage using the aQCMC approach with Iterative QAE~\cite{grinko2021iterative} to estimate physical observable with quadratically less resources.
    }
\end{figure}

\begin{figure}
    \centering
    \includegraphics[width=0.44\textwidth]{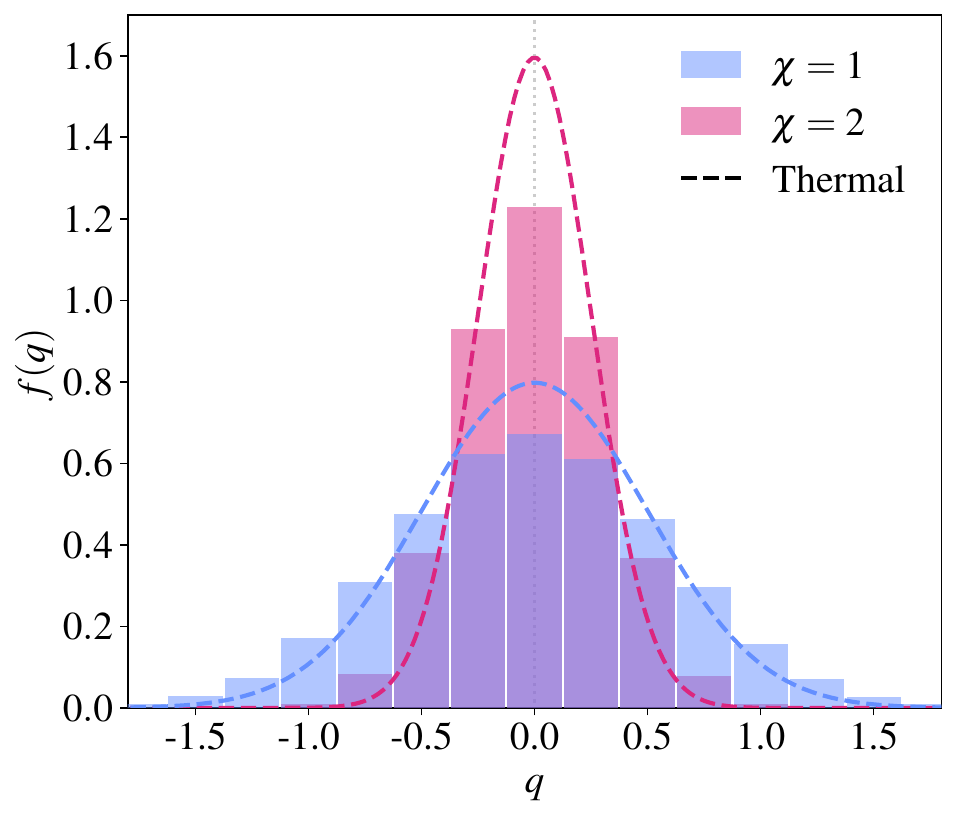}
    \caption{\label{fig:qc_dist} Probability density distribution for quantum simulation with an anisotropic medium at late time $\tilde t = 3$, compared with analytical thermal equilibriums distribution $f_\mathrm{eq}(q)$ in Eq.~(\ref{eq:feq_no_unit}).
    }
\end{figure}

With the theory of heavy quark thermalization and its quantum strategy described above, we present the numerical simulation results of 1D and 2D heavy quark thermalization, including both isotropic and anisotropic mediums. While one can use real quantum computers to simulate the heavy quark thermalization, these devices are still limited today by their short coherence time. Therefore, for the purpose of our investigation, we use the $\mathsf{QASM}$ simulator provided by $\mathsf{Qiskit}$ to mimic ideal quantum devices.
Regarding the physical scales we choose in Eq.~(\ref{eq-langevin_unity}), the heavy quark with a mass $M=1.5$\,GeV in a typical plasma temperature in heavy-ion collisions $T\simeq 300$ - $500$\,MeV gives a range of the variance $\tilde{\sigma}_i^2d\tilde{t}=2Td\tilde{t} / (M\chi_i^2)$ around $2d\tilde{t}/ (5\chi_i^2)$ - $2d\tilde{t}/({3\chi_i^2})$. For simplicity, we may just consider $\tilde{\sigma}_i^2d\tilde{t}\simeq d\tilde{t}/(2\chi_i^2)$.

We first study the heavy quark thermalization using the QCMC approach in a one-dimensional medium, following the quantum circuit in Fig.~\ref{fig:QMCwithReset}. In general, we can simulate the stochastic process with any system size and time step; however, in practice, we are limited by the classical simulation resources. 
Here, for practical purposes, we consider a small system of $n_{\mathcal{S}} = n_{\mathcal{W}} = 4$ qubits and $q\in [-q_{\mathrm{max}}, -q_{\mathrm{max}}) = [-2, 2)$ with momentum resolution $\delta q=0.25$ and time interval $d\tilde{t} = 1/2^d = 0.5$. Together with registers to store the intermediate quantum states, a total of 12 qubits and 8192 shots are used. 
We present the numerical simulation results in Fig.~\ref{fig:qc_1D}. The upper panel of Fig.~\ref{fig:qc_1D} shows a collection of heavy quark momentum $q$ trajectories as an attractor toward thermal expectation. Simulation with two different parameters $\chi=1$, and $\chi=2$ are used. 
Similarly, the lower panel of Fig.~\ref{fig:qc_1D} shows a collection of heavy quark momentum absolute value $|q|$ trajectories.
In addition, we also present the large time momentum distribution at $\tilde{t}=3$ from the simulations with $\chi=1,2$ compared to thermal 1-D distributions $f^{\mathrm{eq}}(q)$ in Eq.~\eqref{eq:feq_no_unit} with $\tilde{\sigma}^2=1/(2\chi^2)=1/2,1/8$. The comparison is shown in Fig.~\ref{fig:qc_dist}. The simulations agree with the expectations with small discrepancies caused by both momentum and time lattice effects.
We also observe that the thermalization with a larger anisotropic parameter $\chi$ leads to thermal distribution with a narrower collection of momentum, characterized by a smaller value of variance $\tilde{\sigma}^2$. 

\begin{figure*}
    \centering
    \subfigure[\label{fig:isotropic} Isotropic medium]{
    \includegraphics[width=0.19\textwidth]{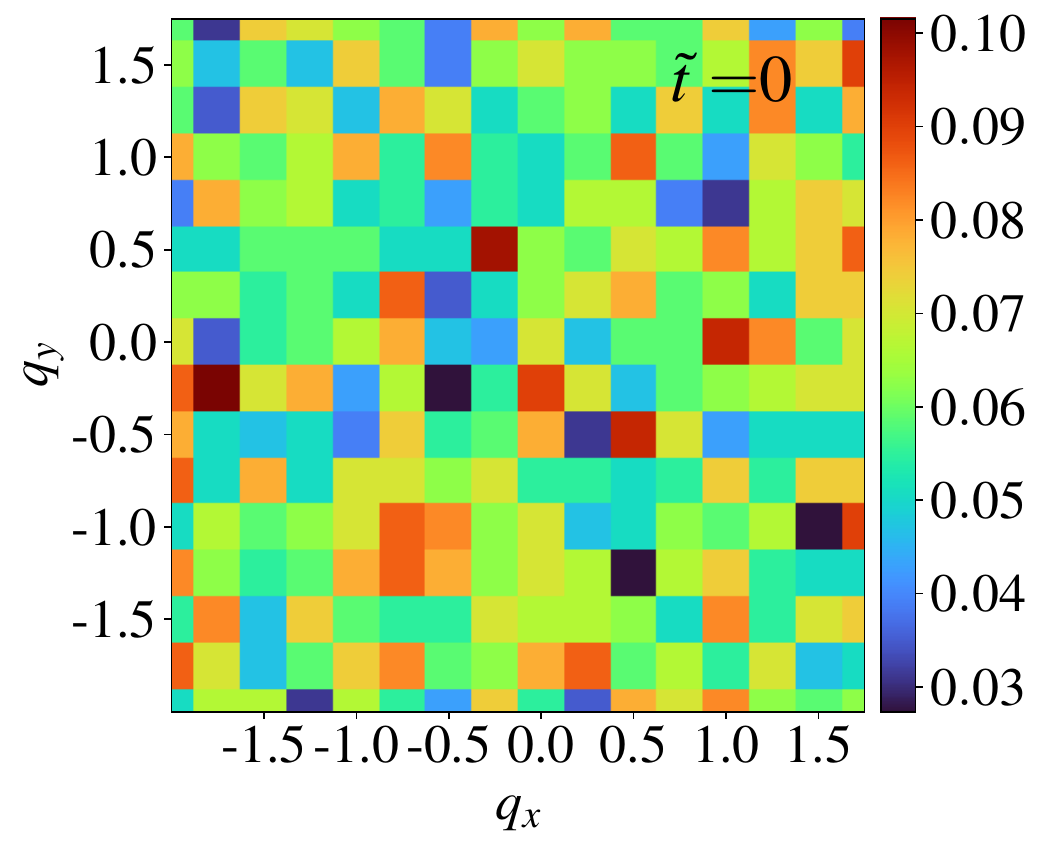}
    \includegraphics[width=0.19\textwidth]{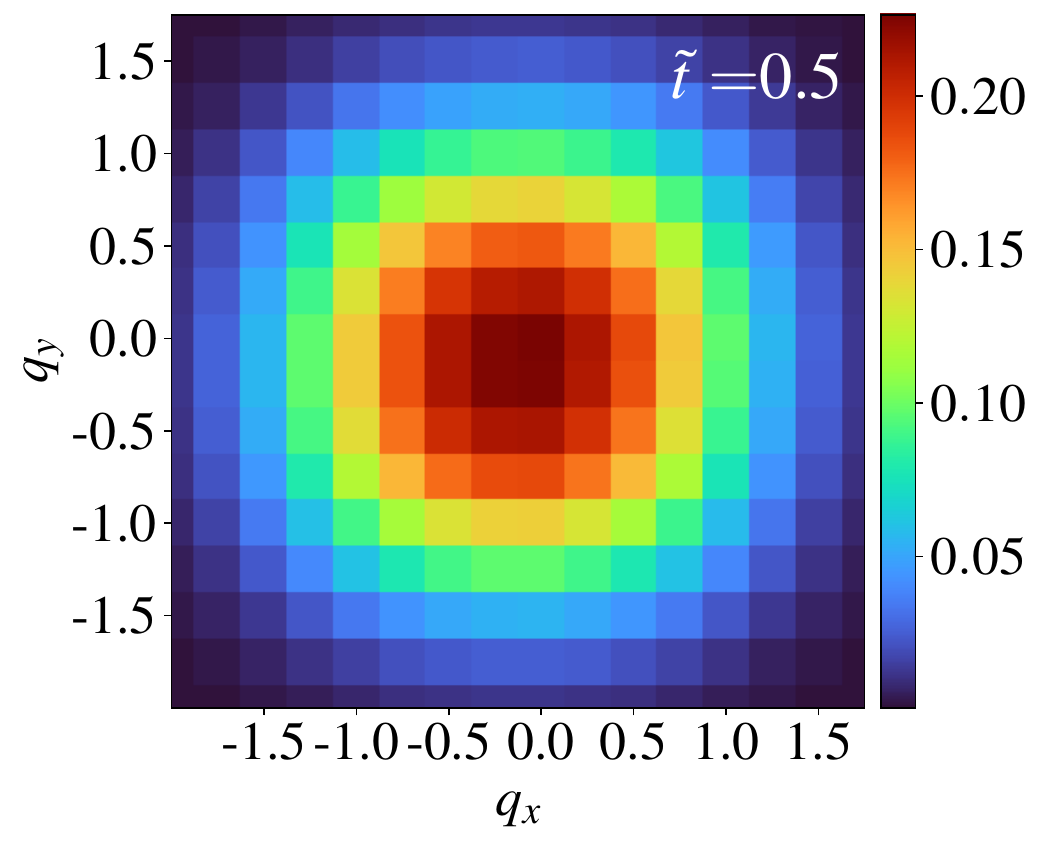}
    \includegraphics[width=0.19\textwidth]{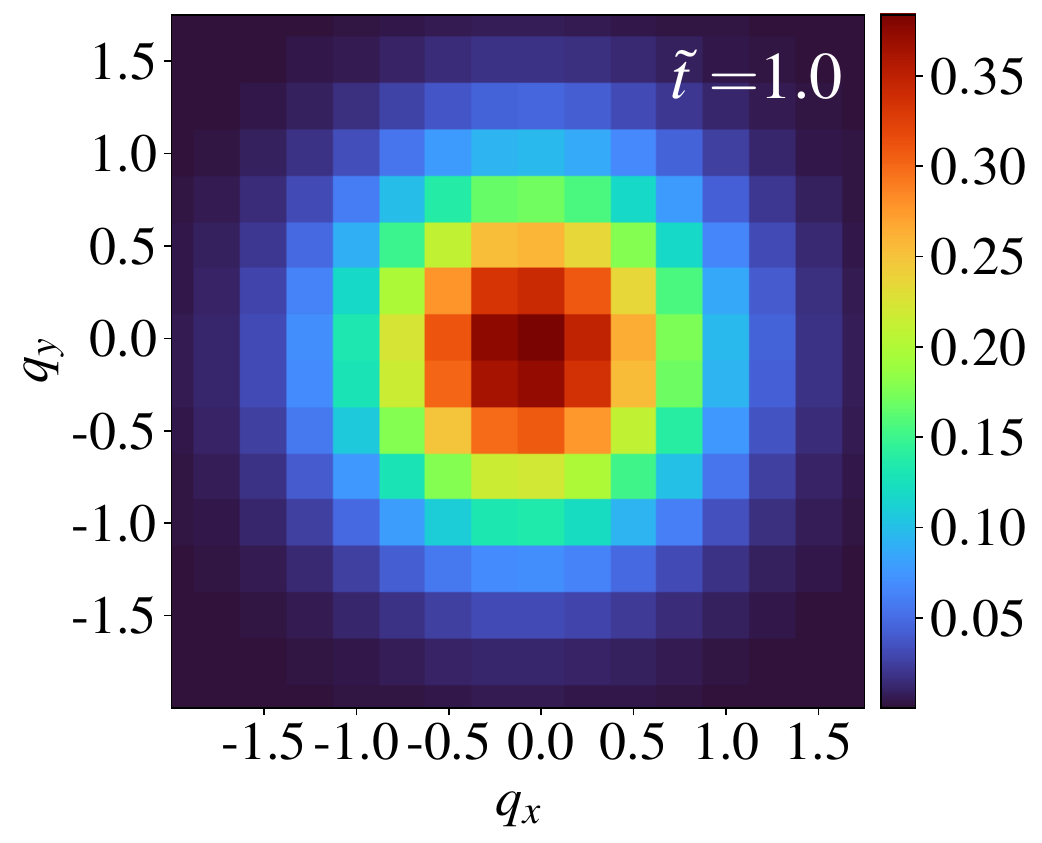}
    \includegraphics[width=0.19\textwidth]{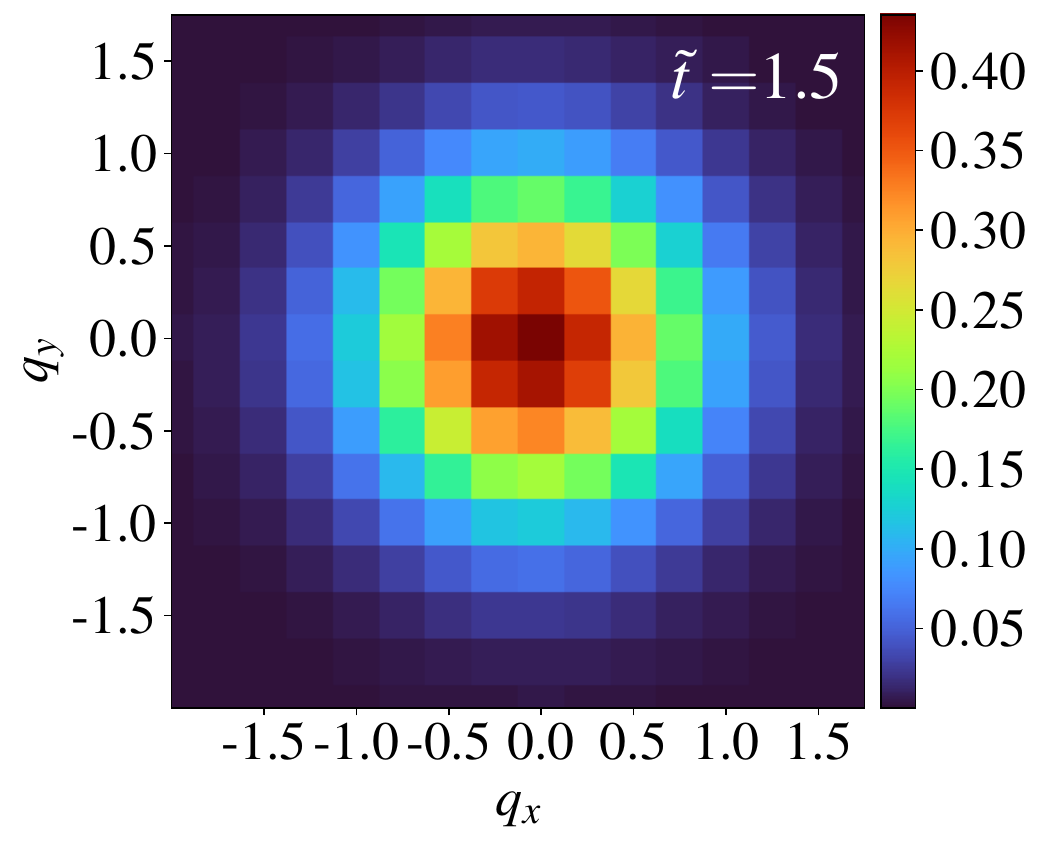}
    \includegraphics[width=0.19\textwidth]{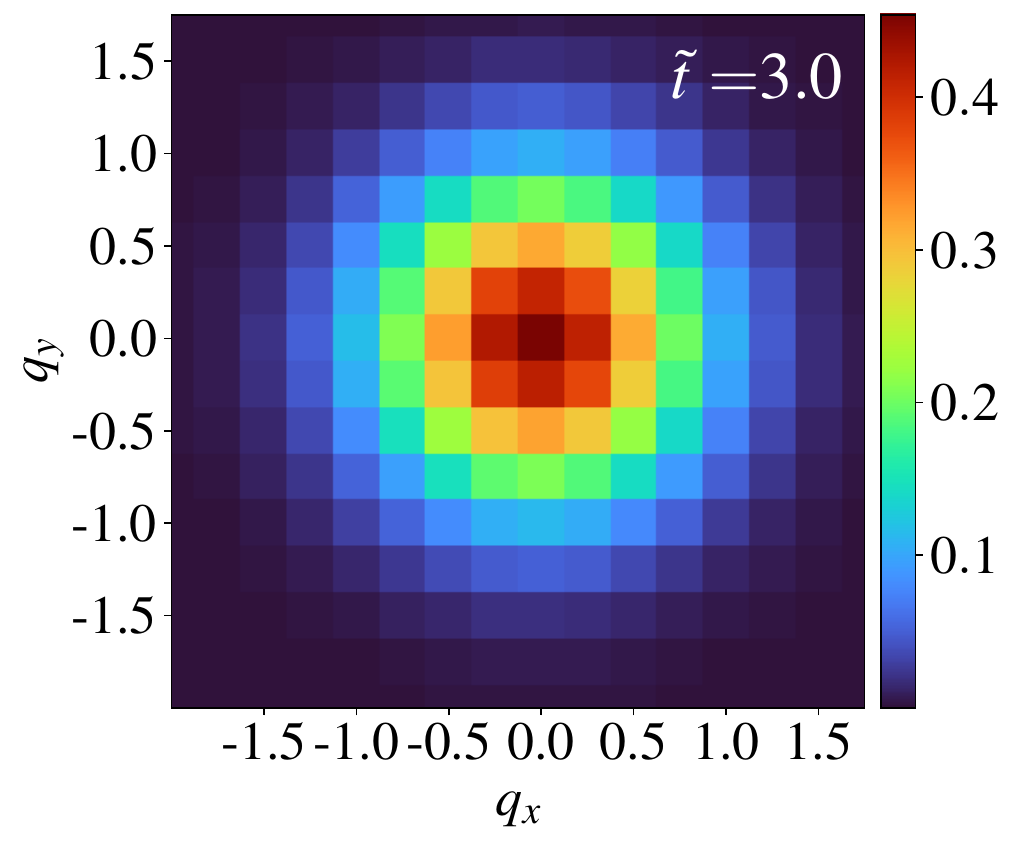}}
    \subfigure[\label{fig:anisotropic} Anisotropic medium]{
    \includegraphics[width=0.19\textwidth]{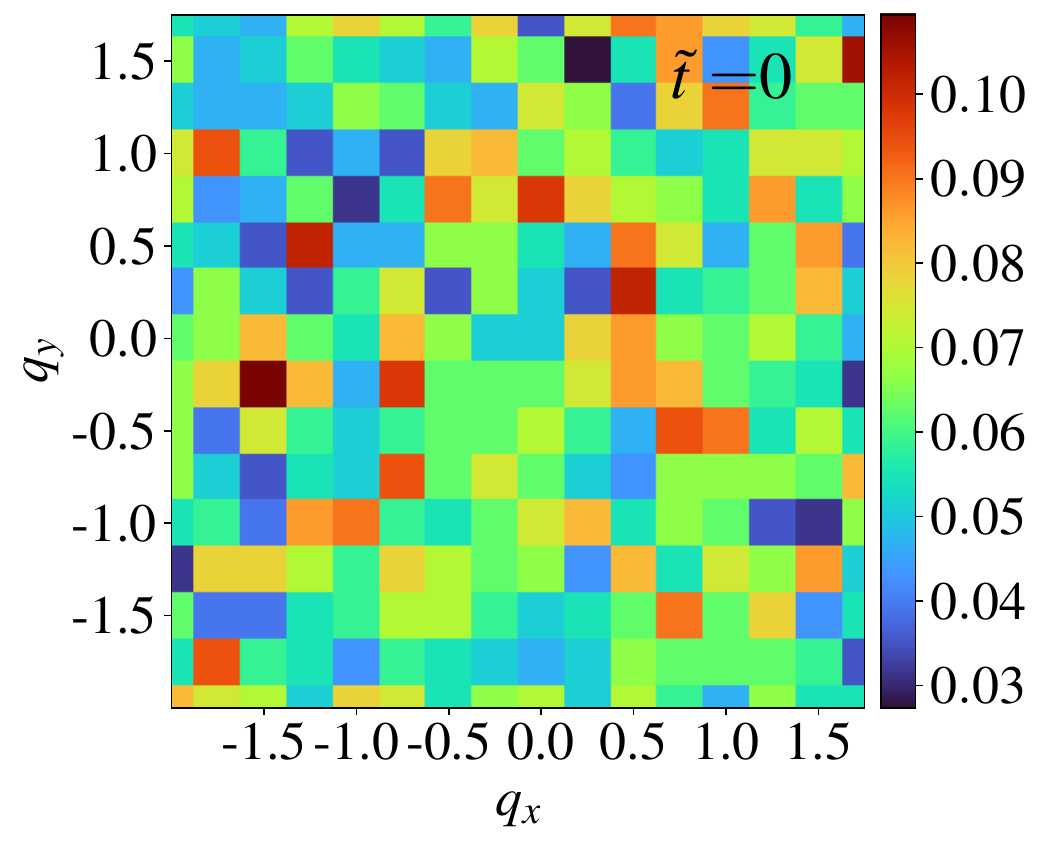}
    \includegraphics[width=0.19\textwidth]{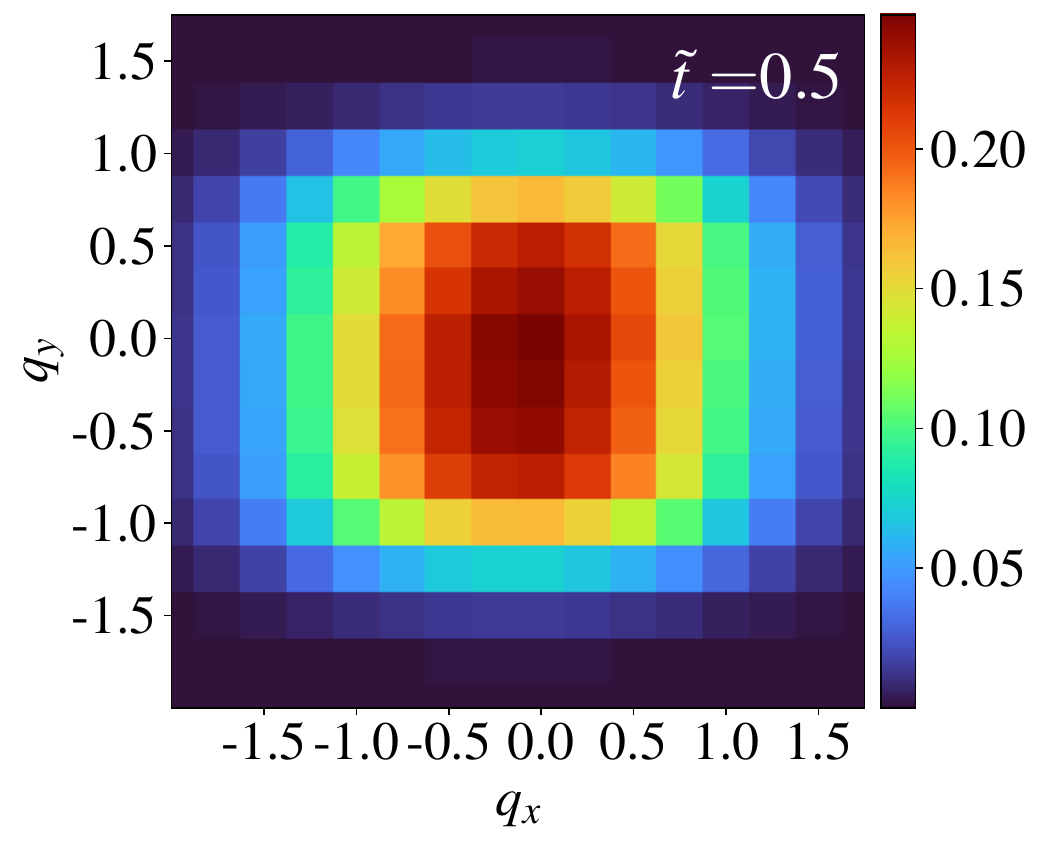}
    \includegraphics[width=0.19\textwidth]{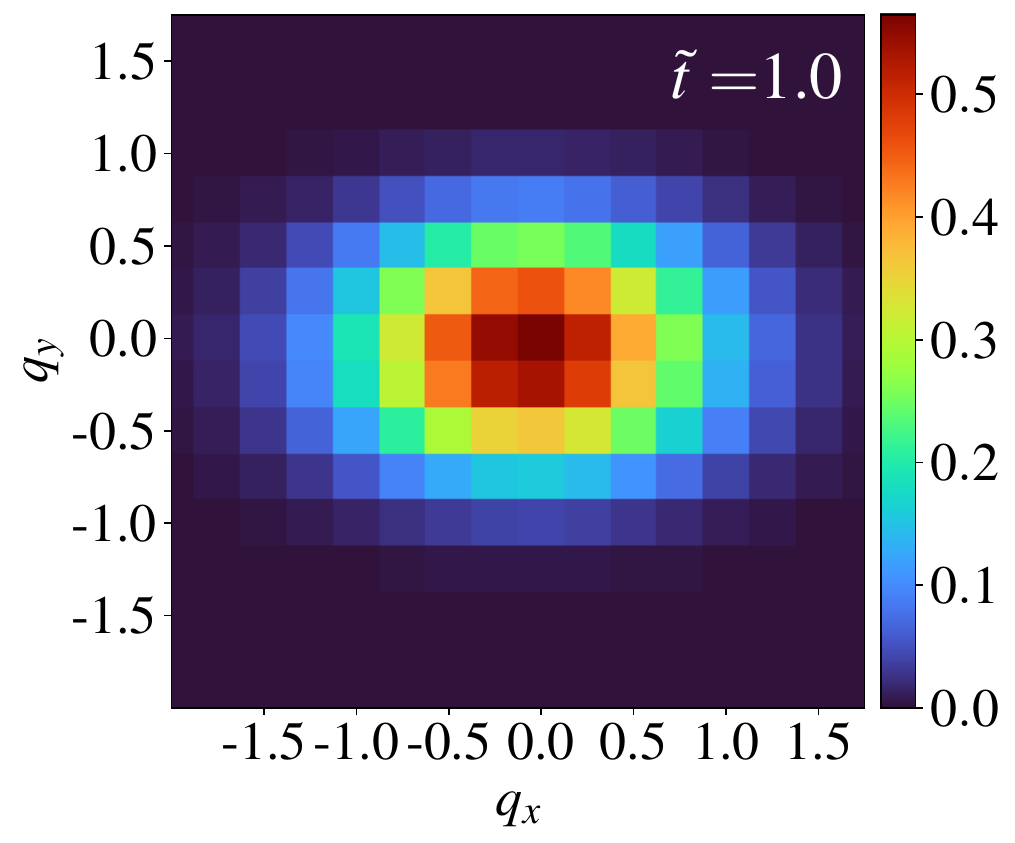}
    \includegraphics[width=0.19\textwidth]{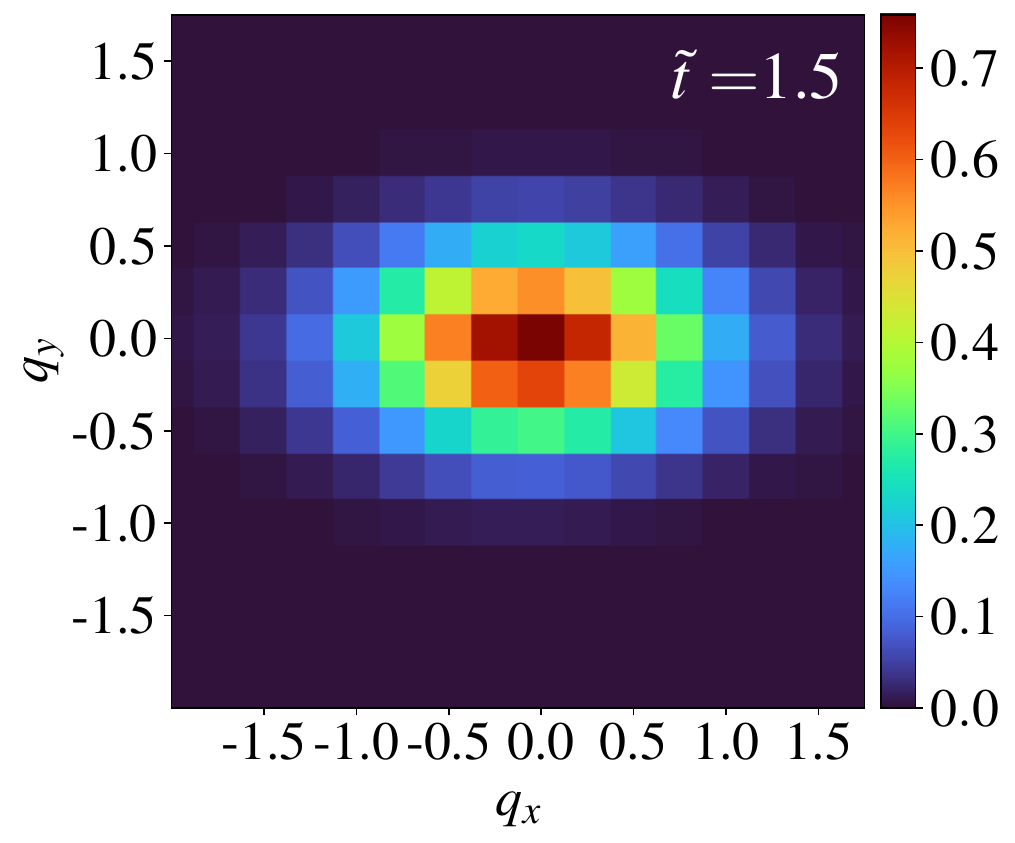}
    \includegraphics[width=0.19\textwidth]{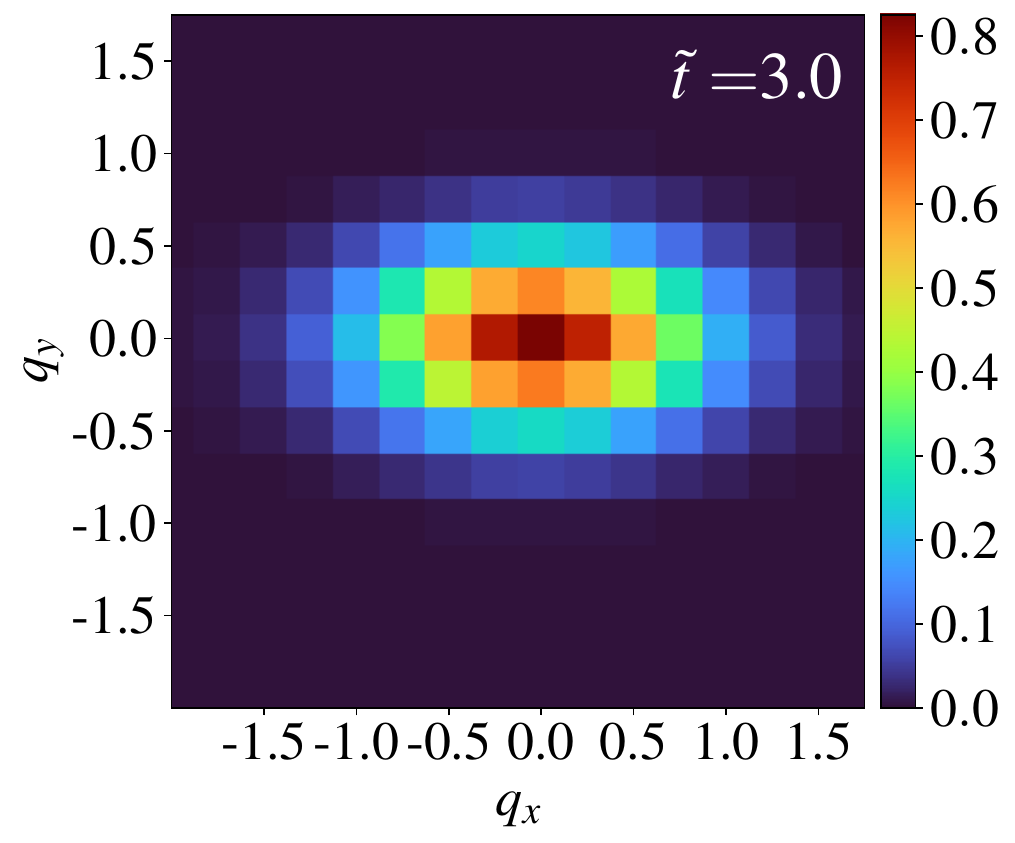}}
    \caption{\label{fig:qc_2D_evo}Momentum distribution with both isotropic and anisotropic mediums for two-dimensional heavy quark thermalization. The vertical axis represents $q_y$; the horizontal axis represents $q_x$. Heat maps at times $\tilde{t} = 0, 0.5, 1.0, 1.5, 3.0$ are presented.}
\end{figure*}

\begin{figure}
    \centering
    \includegraphics[width=0.42\textwidth]{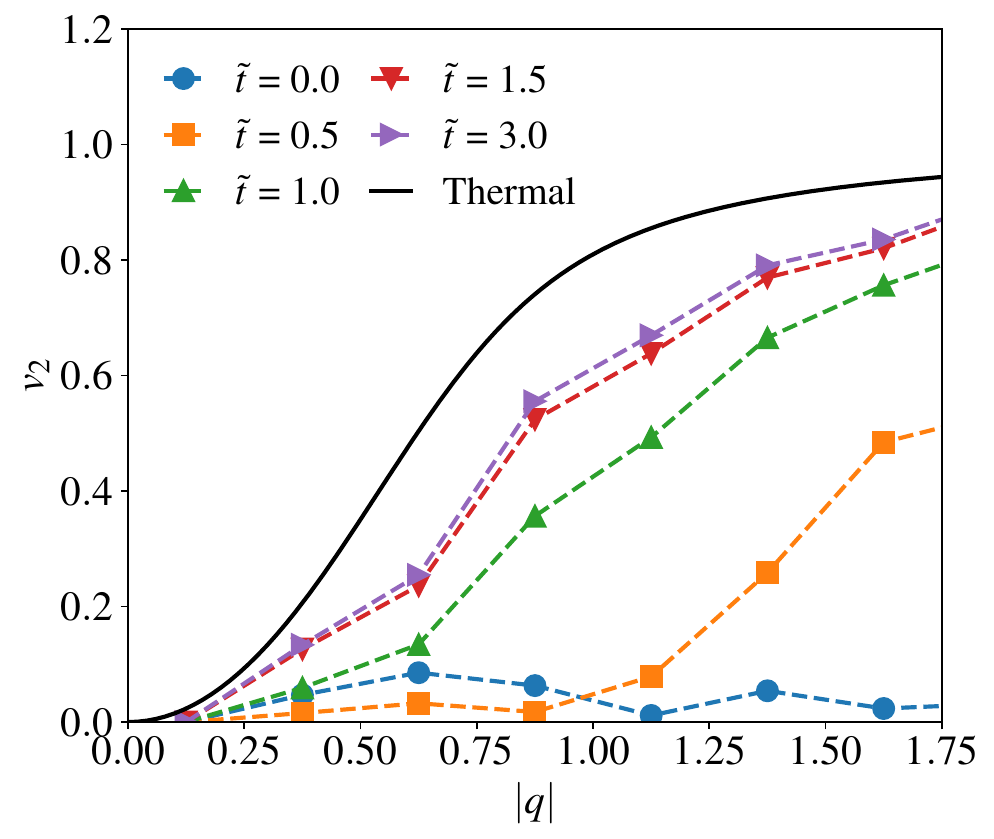}
    \caption{\label{fig:qc_v2} Elliptic flow over time steps using quantum simulation, compared with analytical thermal equilibrium limit in Eq.~\eqref{eq:v2}.
    }
    \label{fig:v2}
\end{figure}

To speed up the QCMC, we implement the QAE to directly extract the expectation $\braket{q}$ and $\braket{|q|}$, following the aQCMC circuit in Fig.~\ref{fig:QMCwithQAE}. In particular, we focus on Iterative QAE (IQAE) for its economical complexity in the calculation~\cite{grinko2021iterative}, and present the simulation results\footnote{Here, we only considered initial $\bar{q}/\delta q=[0,1,2,...,14]$ and omitted $\bar{q}/\delta q=15$ for the simulation to avoid boundary effects due to quantum integer division (see App.~\ref{app:quantum_gates}).}
for the early time steps in Fig.~\ref{fig:qc_1D_QAE}. Without using the {\tt reset} gates, the qubit number grows linearly with the number of time steps, so we simulate to a maximum step of 3, taking up a total of 20 qubits, with an additional ancilla qubit for QAE. We can see that both the aQCMC results agree with the QCMC results within the uncertainty band obtained from 10 simulation batches. Specifically, we used IQAE with an estimation accuracy $\epsilon=0.01$, a confidence level $(1-\alpha)=95\%$, and $N_{\mathrm{shots}}=30$ shots per iteration~\cite{grinko2021iterative}. In total, we used $N_q \approx 1000\pm 240$ shots per each data point for aQCMC\footnote{Fluctuations in the total shots $N_q$ are not statistical, but only due to the iterative process by nature of the IQAE algorithm~\cite{grinko2021iterative}.}, which agrees well with the QCMC results using a large number shots (81920 shots).

To show the quantum advantage using the aQCMC approach with the IQAE, we present the estimation accuracy versus the number of oracle queries $N_q$ in Fig.~\ref{fig:quantum_advantage}. The estimation accuracy is evaluated as the absolute error of the measured momentum to analytical expectation $\epsilon=|\braket{q}-\braket{q}_{\rm analytical}|$, where for convenience, we start with $q_0=0$ and thus $\braket{q}_{\rm analytical}=0$. 
We estimate the absolute error in both direct measurement (DM) and the IQAE using the same total number of shots at $\tilde{t}=0.5$. 
The DM is equivalent to the classical strategy of statistical measurement while the IQAE is its quantum counterpart.
We observe the estimation accuracy $\epsilon$ follow exactly the $\mathcal{O}(1/N_q)$, offering a quadratic boost to classical strategy with $\mathcal{O}(1/\sqrt{N_q})$. Notably, our results are within the theoretical upper bound $N_{q,\mathrm{max}} = (50/\epsilon) \log[(2/\alpha) \log(\pi/(4\epsilon))]$~\cite{grinko2021iterative}.
The same quadratic quantum speedup is also found for the Maximum Likelihood QAE~\cite{suzuki2020amplitude}. 

The quantum simulation strategy introduced in this work, both the QCMC and the aQCMC, can also be applied to two-dimensional heavy quark systems with both isotropic and anisotropic mediums. Here, we consider an isotropic medium with $\chi_x=\chi_y=1$ thus $\tilde{\sigma}^2_{xx}=\tilde{\sigma}^2_{yy}=1/2$, and an anisotropic medium with $\chi_x=1$, $\chi_y=2$ thus $\tilde{\sigma}^2_{xx}=1/2$, $\tilde{\sigma}^2_{yy}=1/8$. 
Due to the large amount of total qubits required for a complete two-dimensional circuit, the simultaneous calculation of the $x,y$ directions in one circuit is not practical with the current hardware. Instead, since the heavy quark dynamics in $x,y$ directions are decoupled with diagonal coefficients, we can simulate $x$ and $y$ directions
separately and pair the events randomly to perform the calculation in two dimensions.
With uniformly distributed heavy quarks as the initial condition, we present the heavy quark thermalization over time in Fig.~\ref{fig:qc_2D_evo}. We can see that different thermalization patterns reflect accordingly for the different medium properties, isotropic and anisotropic. 
In both cases, the collections of heavy quarks reach the thermal distributions provided by Eq.~(\ref{eq:feq_no_unit}).

In the anisotropic medium, one may further evaluate the buildup of the elliptic flow $v_2$, which characterizes the anisotropization due to the medium profile,
\begin{align}\label{eq:v2}
    v_2&=\frac{\int f(q,\cos(\phi),t)\cos( 2\phi) \mathrm{d}\phi}{\int f(q,\cos(\phi),t)\mathrm{d}\phi} \overset{\rm thermal}{=}\frac{I_1(\frac{1}{2q^2}|\frac{1}{\tilde{\sigma}_x^2}-\frac{1}{\tilde{\sigma}_y^2}|)}{I_0(\frac{1}{2q^2}|\frac{1}{\tilde{\sigma}_x^2}-\frac{1}{\tilde{\sigma}_y^2}|)},
\end{align}
where the $v_2$ in the thermal equilibrium is a ratio of modified Bessel functions $I_1(x)$ and $I_0(x)$. In Fig.~\ref{fig:v2}, we calculate $v_2$ using the simulation result. Despite the discrepancy between the simulated $v_2$ at a late stage compared to the analytical thermal limit due to insufficient lattice, we observe a gradual buildup of the $v_2$ for heavy quarks in an anisotropic medium that approaches the limit.

\section{Conclusion and outlook}\label{sec:conclusion}
In this work, for the first time, we present a quantum strategy for stochastically simulated heavy quark thermalization with the QCMC and the aQCMC algorithms on the circuit. 
Specifically, we simulate the heavy quark thermalization with both the QCMC with a longer evolution time step and the aQCMC with a shorter evolution time step but boosted with amplitude estimation.
With these algorithms, we study heavy quark thermalization in both one-dimensional and two-dimensional mediums, as well as isotropic and anisotropic mediums. We show their thermalization patterns and late-time behaviors compared to the analytical expectations. We also calculate the buildup of the elliptic flow for heavy quarks in an anisotropic medium.
Remarkably, with Grover-like quantum amplitude estimation, we can estimate physical observables with quadratically a smaller number of simulation shots, compared to classical Monte Carlo methods, which prepares for quantum advantage in future fault-tolerant simulation.

Notably, the quantum strategies utilized in the work, the QCMC and aQCMC algorithms, are generic for simulating stochastic processes in even broader contents in physics, where the aQCMC has the potential to speed up the calculation in comparison to classical methods.
For instance, with proper modeling of the quark coalescence, this framework can be extended to study quarkonium dissociation and recombination in a medium. Furthermore, real-time dynamics of heavy quark/quarkonium simultaneous production involving many heavy quark/anti-quark pairs will benefit from the quantum boost that significantly reduces the number of samples in such stochastic simulation.
The classical Wiener process may be replaced by a quantum random walk, which eventually gives quantum statistics instead of classical statistics, and a stochastically quantum thermalization of a heavy quark, or spin-chain system might be achieved on a quantum circuit. We leave these for future studies.

\section*{Acknowledgement}
We are grateful to João Barata, Shanshan Cao, Oscar Garcia-Montero, Meijian Li, Tan Luo, Alberto Manzano, Aleksas Mazeliauskas, Carlos A. Salgado, Sören Schlichting, Juan Santos Suárez, Bin Wu, Jianhui Zhang, and Kai Zhou for their helpful and valuable discussions. We acknowledge the use of IBM Quantum services for this work. The views expressed are those of the authors and do not reflect the official policy or position of IBM or the IBM Quantum team. This work is supported by the European Research Council under project ERC-2018-ADG-835105 YoctoLHC; by the Spanish Research State Agency under project PID2020-119632GB-I00; by Xunta de Galicia (Centro singular de investigacion de Galicia accreditation 2019-2022), and by European Union ERDF. WQ is also supported by the Marie Sklodowska-Curie Actions Postdoctoral Fellowships under Grant Agreement No. 101109293.

\bibliography{main_clean.bib}

\appendix

\section{Non-relativistic limit of heavy quark thermalization}\label{app:nr_limit}
In the relativistic case, the Langevin dynamic Eq.~(\ref{eq-langevin}) thermalizes the heavy quark, which results in a Boltzmann distribution $f^{\rm eq}(\vec{x},\vec{p})\propto {\rm exp}(-E(\vec{p})/T)$ with relativistic dispersion relation $E(\vec{p})=\sqrt{\vec{p}^2+M^2}$ upon the coefficients satisfying Einstein's relation
\begin{align}
A(\vec{x},\vec{p},t)p_i
=&\frac{1}{E(\vec{p})}\left(\frac{B_{ij}(\vec{x},\vec{p},t)}{T}-\frac{\partial B_{ij}(\vec{x},\vec{p},t)}{\partial E(\vec{p})}\right)p_j.
\label{eq:einstein_relation}
\end{align}
In the non-relativistic limit, one has a decomposition of the kinetic energy and the mass terms $E(\vec{p})\simeq \vec{p}^2/(2M)+M$, as well as a simplified Einstein's relation $Ap_i=B_{ij}p_j/(MT)$ which results in a Maxwell-Boltzmann distribution $f^{\rm eq}(\vec{p})\propto {\rm exp}(-\vec{p}^2/(2MT))$. 

The realistic diffusion coefficients in multidimensional space are complicated due to non-trivial off-diagonal terms $\sigma_{xy},\sigma_{xz},\sigma_{yz}$.
A set of diagonalized diffusion coefficients $\sigma_{ij}={\rm diag}(\sigma_{xx},\sigma_{yy},\sigma_{zz})$ may be found with a simplied medium profile. In this case, one has diagonalized  coefficient $B_{ij}=\frac{1}{2}{\rm diag}(\sigma_{xx}^2,\sigma_{yy}^2,\sigma_{zz}^2)$ as well. 
However, one might still be interested in an anisotropic medium such that $\sigma_{xx}^2\ne\sigma_{yy}^2\ne\sigma_{zz}^2$, which leads to an anisotropic Langevin equation but with diagonalized terms only
\begin{align}
dp_i&=-A p_i dt + \sigma_{ii} dW_{i},~~~i=x,y,z.
\label{eq-langevin_ani}
\end{align}
With a proper choice of Einstein's relation augmented by a set of anisotropic parameters $\chi_x^2, \chi_y^2, \chi_z^2 \in\mathbb{R^+}$ such that
\begin{align}
A&=\frac{\sigma_{xx}^2\chi_{x}^2}{2MT}=\frac{\sigma_{yy}^2\chi_{y}^2}{2MT}=\frac{\sigma_{zz}^2\chi_{z}^2}{2MT},
\label{eq-einstein-ani}
\end{align}
the Langevin equation Eq.~(\ref{eq-langevin_ani}) approaches a generic anisotropic thermal distribution at the non-relativistic limit, in terms of the momentum $\vec{p}$, heavy quark mass $M$, temperature $T$, and the anisotropic parameters $\chi_{i}^2$
\begin{align}
f^{\rm eq}(\vec{p})&\propto {\rm exp}\bigg[- \frac{\chi_{x}^2p_x^2+\chi_{y}^2p_y^2+\chi_{z}^2p_z^2}{2MT}\bigg].
\end{align}
One may rescale the above equation into dimensionless variables to simplify the discussions and simulations.
Divide the Langevin equation Eq.~(\ref{eq-langevin_ani}) by heavy quark mass $M$ and use Einstein's relation Eq.~(\ref{eq-einstein-ani}), we can reformulate the evolution as
\begin{align}
dq_i=-q_id\tilde{t}+d\tilde{W}_i,~~~i=x,y,z,
\end{align}
with new and dimensionless momentum $q_i=p_i/M$, time $d\tilde{t}=Adt$, and stochastic term $d\tilde{W}_i\sim\mathcal{N}(0,2Td\tilde{t}/(M\chi_i^2))$.

\section{Quantum arithmetic gates}\label{app:quantum_gates}
\begin{figure}[!h]
    \centering
    \raisebox{0.2em}{\subfigure[\label{fig:Qaddxy}]{
    \includegraphics[width=0.22\textwidth]{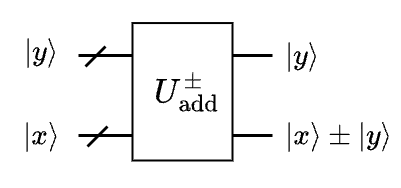}}}
    \subfigure[\label{fig:Qaddxc} ]{
    \includegraphics[width=0.22\textwidth]{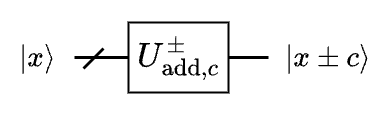}}
    \subfigure[\label{fig:Qshiftxd} ]{
    \includegraphics[width=0.22\textwidth]{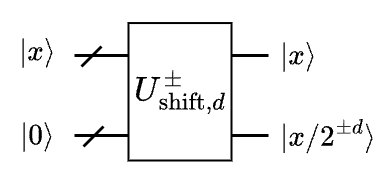}}
    \caption{\label{fig:qc_adder}Schematics of quantum arithmetic gate.}
\end{figure}

We show the basic quantum arithmetic gates used in constructing the quantum circuits for Monte Carlo simulations in Fig.~\ref{fig:qc_adder}, 
\begin{enumerate}[label=(\alph*)]
    \item Quantum adder/subtractor $U^{\pm}_\mathrm{add}\ket{y}\otimes\ket{x} = \ket{y}\otimes(\ket{x} \pm \ket{y})$ on arbitrary states $\ket{x}$ and $\ket{y}$,
    \item Quantum adder/subtractor $U^{\pm}_{\mathrm{add}, c}\ket{x} = \ket{x\pm c}$ on a state $\ket{x}$ and a constant integer $c$,
    \item Quantum multiplier, or bit-shifting unitary $U^{\pm}_{\mathrm{shift},d} \ket{x} = \ket{x/2^{\pm d}}$ on a state $\ket{x}$ and a shift integer $d$.
\end{enumerate}
Note all these three gates follow the integer module $N=2^n$ arithmetics where $n$ is the number of qubits for register $\ket{x}$, and they are responsible for constructing $U_A$ gates in Eq.~
\eqref{eq:UA_gate}. To compensate for float-number quantum multiplication, we find it necessary to prepend the bit-shifting gate with integers and then average. In the case of $d=1$, $\ket{x/2} \equiv \big (U^+_{\mathrm{shift},1}\ket{x} + U^+_{\mathrm{shift},1}\ket{x+1} \big) / 2$.

\section{Quantum amplitude estimation circuits}\label{app:qae_circuit}
We review traditional techniques to perform quantum amplitude estimation (QAE) using quantum phase estimation (QPE)~\cite{nielsen_chuang_2010, brassard2002quantum} with auxiliary qubits and the QPE-free estimation methods without extra qubits~\cite{suzuki2020amplitude, grinko2021iterative, Nakaji_2020}. 
In the simulation of heavy quark thermalization, the system at time $\tilde{t}$ is characterized as a momentum state $\ket{\psi}_n$ stored in a register with a number of $n$ qubits,s
\begin{align}\label{eq:qae}
    \ket{\psi}_n = \sum_{i=0}^{2^n-1}\sqrt{\mathcal{P}(i)}\ket{i}_n, \quad \sum_{i=0}^{2^n-1}\mathcal{P}(i) = 1.
\end{align}
The desired quantum amplitude is then loaded by an oracle $A_F: \ket{i}_n\ket{0} \rightarrow \ket{i}_n\big(\sqrt{1-F(i)}\ket{0} + \sqrt{F(i)}\ket{1} \big)$ acting on the state and an ancilla qubit,
\begin{align}
    A_F \ket{\psi}_n\ket{0} &= \sum_{i=0}^{2^n-1}\sqrt{1-F(i)}\sqrt{\mathcal{P}(i)}\ket{i}_n\ket{0} \nonumber \\
    &+ \sum_{i=0}^{2^n-1}\sqrt{F(i)}\sqrt{\mathcal{P}(i)}\ket{i}_n\ket{1} \nonumber\\
    &= \sqrt{1-a}\sum_{i=0}^{2^n-1}\frac{\sqrt{(1-F(i))\mathcal{P}(i)}}{\sqrt{1-a}}\ket{i}_n\ket{0} \nonumber \\
    &+ \sqrt{a}\sum_{i=0}^{2^n-1}\frac{\sqrt{F(i)\mathcal{P}(i)}}{\sqrt{a}}\ket{i}_n\ket{1} \nonumber\\
    &= \sqrt{1-a}\ket{\psi_0}_n\ket{0} + \sqrt{a}\ket{\psi_1}_n\ket{1} \nonumber\\
    &= \sqrt{1-a}\ket{\psi^*_0}_{n+1} + \sqrt{a}\ket{\psi^*_1}_{n+1} \nonumber\\
    &= \ket{\psi^*}_{n+1}
\end{align}
where $a=\sum_{i=0}^{2^n-1} \mathcal{P}(i) F(i) = \braket{F}$ is the expectation value of certain physics quantity $F$ that we are interested in, and new basis $\ket{\psi^*_0}_{n+1}$ and $\ket{\psi^*_1}_{n+1}$ are used. Notably, although the function $F$ is not restricted to a domain $[0, 2^n-1]$ and an image of $[0, 1]$, one can always rescale the target $F$ to be within $[0, 1]$, by applying affine transformation that preserves collinearity,.

\begin{figure}
    \centering
    \raisebox{0.2em}{\subfigure[\label{fig:qae_qpe} QPE circuit, with inverse quantum Fourier transform $\mathcal{F}^{-1}$]{
    \includegraphics[width=0.40\textwidth]{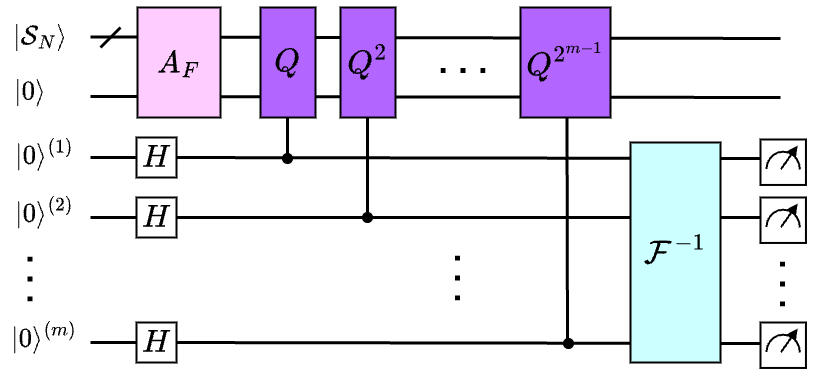}}}
    \subfigure[\label{fig:qae_qpefree} QPE-free circuit ]{
    \includegraphics[width=0.30\textwidth]{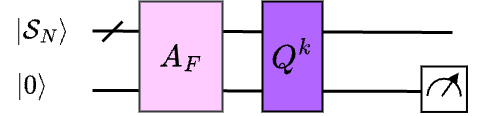}}    
    \caption{\label{fig:qae_circuit}Quantum amplitude estimation circuits.}
\end{figure}

To estimate this $a$, Grover's operator is defined as $\mathcal{Q}=A_F S_0 A_F^{\dagger}S_{\psi_1}$ where the sign-flipping operators are $S_0=I-2\ket{0}_{n+1}\bra{0}_{n+1}$ and $S_{\psi_1}=I-2\ket{\psi_1^*}_{n+1}\bra{\psi_1^*}_{n+1}$. We can also rewrite Grover's operator as applications of two unitaries, $Q \sim U_{\psi^*} U_{\psi^*_0}$, in a similar way as Grover's search, such that
\begin{align}
Q&=(-R_{\psi^*})(-R_{\psi^*_0})=R_{\psi^*}R_{\psi^*_0} \nonumber\\
&=\big(2\ket{\psi^*}_{n+1}\bra{\psi^*}_{n+1} - I_{n+1}\big )R_{\psi^*_0} \nonumber\\
&=\underbrace{A_F \big(2\ket{\psi}_n\ket{0}\bra{0}\bra{\psi}_n - I_{n+1}\big ) A_F^{\dagger}}_{\mathrm{Diffusion}}\underbrace{R_{\psi^*_0}}_{\mathrm{Phase}}
\end{align} 
where the first part $R_{\psi_0^*} = 2\ket{\psi_0^*}\bra{\psi_0^*} - I$ is the phase/reflection oracle and the second part is the diffusion oracle. With a geometrically increasing powers of $\mathcal{Q}^k$ on $m$ ancilla qubits, one can amplify and estimate the desired amplitude $a$ using the QPE circuit (Fig.~\ref{fig:qae_qpe}) as $\bar{a} = y \pi / (2^m)$ for $y\in\{0, 1,\cdots, 2^m-1\}$ on the measured qubits. The estimation of $\bar{a}$ has an error $\epsilon = |a - \bar{a}| = \mathcal{O}(\sqrt{a(1-a)}/(2^m))$\cite{grinko2021iterative}.

The amplitude can also be estimated by applying $\mathcal{Q}^k A_F$ operations, without the need for ancilla qubits. Let $a=\sin^2(\theta_a)$ and we observe that
\begin{align}
\nonumber
\mathcal{Q}^kA_F\ket{\psi}_n\ket{0}&=\cos\big((2k+1)\theta_a\big)\ket{\psi_0}\ket{0}\\
&+\sin\big((2k+1)\theta_a\big)\ket{\psi_1}\ket{1},
\end{align}
where the probability of measuring $\ket{1}$ gives $\sin^2\big((2k+1)\theta_a\big)$.
By selecting different values of $k$ and combining their outcomes, one could also find the estimated value of $a$ with the same estimation error as the QPE. The specific choice of $k$ varies in each algorithm~\cite{Nakaji_2020, grinko2021iterative, suzuki2020amplitude, Wie:2019, Aaronson:2020}, and we provide a general circuit in Fig.~\ref{fig:qae_qpefree} for those QPE-free approaches.

\end{document}